\documentstyle[aps,epsf,rotate,preprint]{revtex}

\begin{document}
\draft

\title{Inverse Cubic Law for the Distribution of Stock
Price Variations}

\author{Parameswaran Gopikrishnan, Martin Meyer, Lu\'{\i}s~A.~Nunes
Amaral,\\ and H. Eugene Stanley}

\address{Center for Polymer Studies and Department of Physics,\\
Boston University
Boston, MA 02215, U.S.A.}

\date{\today}

\maketitle

\begin{abstract} 
The probability distribution of stock price changes is studied by
analyzing  a database (the Trades and Quotes Database)  documenting
every trade for all stocks in  three major US stock markets, for the
two year period Jan 1994 -- Dec 1995. A sample of 40 million data points
is extracted, which is substantially larger than studied hitherto. We
find an asymptotic power-law behavior for the cumulative distribution
with an exponent $\alpha\approx 3$, well outside the Levy
regime $(0<\alpha<2)$. 
\end{abstract}

\bigskip\noindent

The asymptotic behavior of the increment distribution
of economic indices has long been a topic of avid interest 
\cite{Bachelier00,Mandelbrot63,Mantegna95,Ghasghaie96,Pagan96,Cont97b}.
Conclusive empirical results are, however, difficult to obtain, since
the asymptotic behavior can be obtained only by a proper sampling of
the tails, which requires a huge quantity of data. Here, we analyze a
database documenting each and every trade in the three major US stock
markets, the New York Stock Exchange (NYSE), the American Stock
Exchange (AMEX), and the National Association of Securities Dealers
Automated Quotation (NASDAQ) for the entire 2-year period Jan.~1994 to
Dec.~1995 \cite{TAQ}. We thereby extract a sample of approximately
$4\times10^7$  data points, which is much larger than studied
hitherto. 

We form 1000 time series $S_i(t)$, where $S_i$
is the market price of company $i$  (i.e.~the share price multiplied with
the number of outstanding shares), $i=1\dots1000$ is the rank of the
company according to its market price on Jan.~1, 1994. The basic 
quantity of our study is the relative price change,
\begin{equation}
G_i(t)\equiv \ln S_i(t+\Delta t) -\ln S_i(t) \simeq{S_i(t+\Delta t)-S_i(t)\over
S_i(t)}\;,
\label{eq_defG}
\end{equation}
where the time lag is $\Delta t=5\,$min. We normalize the increments,
\begin{equation}
g_i(t)\equiv \left[G_i(t) -\langle G_i(t)\rangle\right]/v_i\;,
\label{eq_defv}
\end{equation}
where the volatility $v_i\equiv\sqrt{\langle (G_i(t) -\langle
G_i(t)\rangle)^2\rangle}$ of company $i$ is measured by the standard
deviation, and $\langle\dots\rangle$ is a time average \cite{std_comm}.

\begin{figure}

\begin{center}
\begin{minipage}{15cm}
\epsfysize=13cm
\rotate[r]{\epsffile{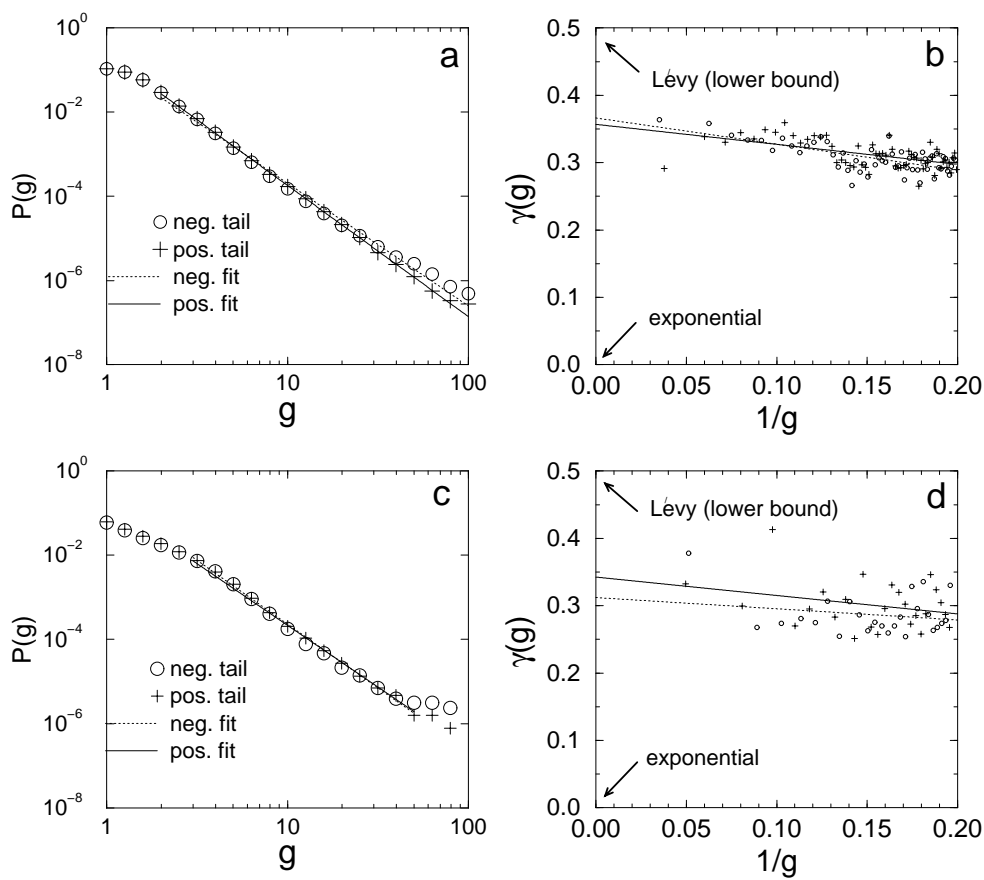}}

\caption{(a) Log-log plot of the cumulative probability distribution
$P(g)$ of the normalized price increments $g_i(t)$. The lines in (a)
are power law fits to the data over the range $2<g<100$.  (b) the
inverse local slope of $P(g)$, $\gamma(g)\equiv -\left(d\log P(g)/d\log
g\right)^{-1}$ as a function of $1/g$ for the negative $(\circ)$ and
the  positive $(+)$ tail respectively. We obtain an estimator for
$\gamma(g)$, by sorting the normalized increments by their size,
$g^{(1)}>g^{(2)}>...>g^{(N)}$. The cumulative density can then be
written as $P(g^{(k)})=k/N$, and we obtain for the local slope
$\gamma(g^{(k)})=-\log(g^{(k+1)}/g^{(k)})/\log(
P(g^{(k+1)}/P(g^{(k)}))\simeq k(\log(g^{(k+1)})-\log(g^{(k)}))$. Each
data point shown in b is an average over 1000 increments $g^{(k)}$, and
the lines are linear regression fits to the data. Note that the average
$m^{-1}\sum_{k=1}^m \gamma(g^{(k)})$ over all events used would be
identical to the estimator for the asymptotic slope proposed by Hill
\protect\cite{Hill75}. (c) Same as (a) for the 1 min S\&P500
increments. The regression lines yield $\alpha=2.93$ and $\alpha=3.02$
for the positive and negative tails respectively. (d) Same as (b) for
the 1 min S\&P500 increments, except that the number of increments per
data point is 100.} 
\label{fig_slope}  
\end{minipage}
\end{center}
\end{figure}

We obtain about 20,000 normalized increments $g_i(t)$ per company per
year, which gives about $4\times10^7$ increments for the 1000 largest
companies in the time period studied. Figure \ref{fig_slope}a shows the
cumulative probability distribution, i.e.~the probability for an
increment larger or equal to a threshold $g$, $P(g)\equiv P\{g_i(t)\ge
g\}$, as a function of $g$. The data are well fit by a power law  
\begin{equation}
P(g)\sim g^{-\alpha}
\end{equation} 
with exponents $\alpha=3.1\pm0.03$ (positive tail) and
$\alpha=2.84\pm0.12$ (negative tail) from two to hundred standard
deviations.

In order to test this result, we calculate the inverse of the local
logarithmic slope of $P(g)$,  $\gamma^{-1}(g)\equiv-d\log P(g)/d\log g$
\cite{Hill75}. We estimate the asymptotic slope $\alpha$ by
extrapolating $\gamma$ as a function of $1/g$ to $0$. Figure
\ref{fig_slope}b shows the results for the negative and positive tail
respectively, each using all increments larger than 5 standard
deviations. Extrapolation of the linear regression lines yield
$\alpha=2.84\pm0.12$ for the positive and $\alpha=2.73\pm0.13$ for the
negative tail. We test this method by analyzing two surrogate data sets
with known asymptotic behavior, (a) an independent random variable with
$P(x)=(1+x)^{-3}$ and  (b) an independent random variable with
$P(x)=\exp(-x)$. The method yields the correct results 3 and
$\infty$ respectively. 

To test the robustness of the inverse cubic law $\alpha\approx3$, we
perform several additional calculations: $(i)$ we change the time
increment in steps of 5$\,$min up to 120$\,$min, $(ii)$ we analyze the
S\&P500 index over the 13y-period Jan.~'84 -- Dec.~'96 using the same
methods as above (Fig.~\ref{fig_slope}c and d), $(iii)$ we replace
definition of the volatility by other measures, such as
$v_i\equiv\langle |G_i(t) -\langle G_i(t)\rangle|\rangle$. The results
are all consistent with $\alpha=3$. These extensions will be discussed
in detail elsewhere \cite{Gopi98}.

To put these results in the context of previous work, we recall that
proposals for $P(g)$ have included $(i)$ a Gaussian distribution
\cite{Bachelier00}, $(ii)$ a L\'evy distribution
\cite{Mandelbrot63,Pareto1897,Levi37}, and $(iii)$ a truncated L\'evy
distribution, where the tails become ``approximately exponential''
\cite{Mantegna95}. The inverse cubic law differs from all three
proposals: Unlike $(i)$ and $(iii)$, it has diverging higher moments
(larger than 3), and unlike $(i)$ and $(ii)$ it is not a stable
distribution.

{\bf Acknowledgements:} We thank X.~Gabaix, S.~Havlin, Y.~Liu,
R.~Mantegna, C.K.~Peng and D.~Stauffer for helpful discussions, and DFG
and NSF for financial support.

\bigskip



\begin{references} 
\bibitem{Bachelier00} L. Bachelier, Ann. Sci. \'Ecole Norm. Sup.
{\bf 3}, 21 (1900).

\bibitem{Mandelbrot63} B.~B. Mandelbrot,  J. Business {\bf 36},
294 (1963).

\bibitem{Mantegna95}  R.~N.~Mantegna and H.~E.~Stanley,  Nature {\bf
376}, 46 (1995).

\bibitem{Ghasghaie96} S.~Ghashgaie,  W.~Breymann, J.~Peinke, P.~Talkner,
and Y.~Dodge, Nature {\bf 381}, 767 (1996). 

\bibitem{Pagan96} A.~Pagan, J. Empirical Finance {\bf 3},  
15 (1996), and references therein.

\bibitem{Cont97b} J.P.~Bouchaud and M.~Potters, {\it Theorie des
Risques Financieres}, (Alea-Saclay, Eyrolles 1998); R.~Cont,
Europ. Phys. J. B (1998), in press. (cond-mat/9705075).

\bibitem{TAQ} {\it The Trades and Quotes Database}, 24 CD-ROM for
'94-'95, published by the New York Stock Exchange.

\bibitem{std_comm} In order to obtain uncorrelated estimators for the
volatility and the price increment at time $t$, the time average
extends only over time steps $t'\ne t$.

\bibitem{Hill75} B.~M.~Hill, Ann. Stat. {\bf3}, 1163 (1975). 

\bibitem{Gopi98} P.~Gopikrishnan, M.~Meyer, L.~A.~N.~Amaral, and
H.~E.~Stanley, to be published.

\bibitem{Pareto1897} V.~Pareto, {\it Cours d'\'Economie Politique}
(Lausanne and Paris, 1897).

\bibitem{Levi37} P.~L\'evy, {\it Th\'eorie de l'addition des variables
al\'eatoires} (Gauthier-Villars, Paris, 1937).

\end{references}
\end{document}